\documentclass{article}
\usepackage{color}
\usepackage{graphicx}
\usepackage{amsmath, amsthm}
\usepackage{mathtools}
\usepackage{natbib}
\usepackage{url}
\RequirePackage[colorlinks,citecolor=blue,urlcolor=blue]{hyperref}
\usepackage{xcolor}
\usepackage{chngcntr}
\usepackage{subfig}
\usepackage{hyperref}
\usepackage{algorithm}
\usepackage[noend]{algpseudocode}
\usepackage{setspace}
\newtheorem{theorem}{Theorem}

\usepackage{xfrac}
\definecolor{forest}{rgb}{0.133,0.545,0.133}
\usepackage{multirow}
\usepackage{amsfonts}
\newtheorem{lemma}{Lemma}

\usepackage{enumitem}
\usepackage{caption}
\usepackage{array}
\usepackage{makecell}

\evensidemargin=\oddsidemargin
\usepackage{etoolbox}

\newif\ifabbreviation
\pretocmd{\thebibliography}{\abbreviationfalse}{}{}
\AtBeginDocument{\abbreviationtrue}

\begin{document}
	\newcommand{\bb}{\boldsymbol{\beta}}

	\title{Design of Bayesian Clinical Trials with Clustered Data}


	\author{Luke Hagar\footnote{Luke Hagar is the corresponding author and may be contacted at \url{l.hagar@uq.edu.au}.} \hspace{35pt} Shirin Golchi$^{\dagger}$ \bigskip \\ 
 $^*$\textit{Clinical Trials Capability, The University of Queensland} \\ $^{\dagger}$\textit{Department of Epidemiology, Biostatistics \& Occupational Health, McGill University}}

	\date{}

	\maketitle

	\begin{abstract}

In the design of clinical trials, it is essential to assess the design operating characteristics (e.g., power and the type I error rate). Common practice for the evaluation of operating characteristics in Bayesian clinical trials relies on estimating the sampling distribution of posterior summaries via Monte Carlo simulation. It is computationally intensive to repeat this estimation process for each design configuration considered, particularly for clustered data that are analyzed using complex, high-dimensional models. In this paper, we propose an efficient method to assess operating characteristics and determine sample sizes for Bayesian trials with clustered data. We prove theoretical results that enable posterior probabilities to be modeled as a function of the number of clusters. Using these functions, we assess operating characteristics at a range of sample sizes given simulations conducted at only two cluster counts. These theoretical results are also leveraged to quantify the impact of simulation variability on our sample size recommendations. The applicability of our methodology is illustrated using an example cluster-randomized Bayesian clinical trial.

		\bigskip

	\end{abstract}

	\maketitle

	\baselineskip=19.5pt



\section{Introduction}\label{sec:intro}

Bayesian methods for data-driven decision making have become increasingly popular in the design and analysis of clinical trials. Bayesian inference is particularly useful in trials with clustered data given the challenges with numerical stability that can arise in frequentist generalized linear mixed modeling \citep{breslow2004whither,capanu2013assessment}. Decision making in Bayesian clinical trials is often based on posterior and posterior predictive probabilities \citep{spiegelhalter2004bayesian, berry2010bayesian}. In many cases, despite the use of Bayesian analysis and decision criteria, grant and regulatory agencies require the design to be assessed with respect to frequentist operating characteristics \citep{fda2019adaptive}. Certain operating characteristics, such as power and the type I error rate, are considered for most Bayesian trials. In Bayesian trials, sampling distributions of posterior summaries must be accurately estimated to correctly assess the frequentist operating characteristics. Sample size determination (SSD) procedures rely on accurate estimates of these operating characteristics. 

Wang and Gelfand \citep{wang2002simulation} proposed a general framework for Bayesian SSD in which Monte Carlo simulation is used to accurately estimate frequentist operating characteristics of Bayesian designs. This computational approach estimates sampling distributions by simulating many repetitions of a trial with a particular set of design, analysis, and decision parameters. In comprehensive trial design, these numerical studies can involve many scenarios, each of which is defined by a different combination of design, analysis, and decision parameters. Design parameters may include sample sizes; analysis parameters may include the effect size for the estimand of interest and values for nuisance parameters; and decision parameters may include decision thresholds. These simulations often borrow from the frequentist framework for power analysis in that each scenario and its corresponding sampling distribution are defined given fixed choices for the analysis model parameters. Alternative Bayesian SSD approaches in which sampling distributions of posterior summaries are marginalized with respect to a \emph{design prior} that incorporates uncertainty in the model parameters have also been proposed \citep{o2005assurance, de2007using,gubbiotti2011bayesian}. 

The U.S. Food and Drug Administration (FDA) recommends using at least $10^4$ simulation repetitions to estimate the sampling distribution for each scenario considered \citep{fda2019adaptive}. Each simulation repetition for each scenario typically involves computational posterior approximation. The computational burden associated with posterior approximation is substantial for high-dimensional models, including those for clustered data that use random effects to model within-cluster dependence. Clustered data are common in clinical settings (e.g., longitudinal studies and cluster-randomized trials). 

In cluster-randomized trials, cluster-level or individual-level estimands may be of interest depending on the research objectives \citep{Kahan2023estimands}. This paper is concerned with individual-level estimands. In trials where adjusted generalized linear models are employed for analysis, further computing resources are required to construct marginal estimands via Bayesian G-computation \citep{kalton1968standardization, keil2018bayesian, daniel2021making, willard2024covariate}. These estimands are obtained by marginalizing samples from the posterior distribution with respect to the distribution of additional covariates or random effects. Drawing inference based on marginal estimands is important when population average treatment effects are of interest but adjusted analyses are employed \citep{willard2024covariate}. 

It is important to develop methodology that ensures numerical studies efficiently inform operating characteristic assessment and Bayesian SSD. Streamlined design methods reduce costs associated with trial design and enhance timely communication between the stakeholders of a clinical trial. Various strategies have recently been proposed to reduce the computational burden associated with estimating operating characteristics for Bayesian designs. \citet{golchi2022estimating} proposed a modeling approach to estimate sampling distributions of posterior summaries and trial design operating characteristics using Gaussian processes; \citet{golchi2023estimating} presented an alternative modeling method that augmented simulations with asymptotic theory. \citet{hagar2023fast} approximated power curves using segments of the relevant sampling distributions. 

For a given simulation scenario, \citet{hagar2024scalable} proposed a method to assess operating characteristics  across the sample size space using estimates of the sampling distribution at only two sample sizes. In that work, the quantiles of the sampling distribution of the logits of posterior probabilities were shown to depend linearly on the sample size. Hence, numerical studies at only two sample sizes are required to infer the sampling distribution at new sample sizes. However, that contribution and the other aforementioned work focus on Bayesian studies with independent observations. Since decisions in many trials are based on dependent data, we build upon the method from \citet{hagar2024scalable} to facilitate the efficient assessment of operating characteristics in designs with clustered data. Our proposed methods are simple to implement and promote an economical framework for simulation-based Bayesian SSD that is suitable for broad use with clinical trials.

The remainder of this article is structured as follows. In Section \ref{sec:ex}, we describe a Bayesian trial where households are randomized to preventative tuberculosis treatments. This illustrative example contextualizes our paper's methodological development. We introduce background information and notation required to describe our methods in Section \ref{sec:methods}. In Section \ref{sec:proxy}, we prove new theoretical results about a proxy to the sampling distribution of posterior probabilities based on clustered data. In Section \ref{sec:power}, we adapt these theoretical results to develop a procedure for Bayesian SSD that requires estimation of the sampling distribution of posterior probabilities at only two cluster counts. In Section \ref{sec:starlet}, we conduct numerical studies to assess the performance of the proposed SSD method for our illustrative example. Section \ref{sec:ext} discusses extensions to this work for adaptive trials and trials with multiple endpoints. We conclude with a discussion in Section \ref{sec:disc}.

\section{Example, Data, and Estimands}\label{sec:ex}

In this section, we introduce general notation for clustered data in clinical trials. This structure for the data is contextualized using an illustrative example. This example draws inspiration from the Shorter and Safer Treatment Regimens for Latent Tuberculosis (SSTARLET) trial \citep{menzies2024SSTARLET}. SSTARLET aims to establish that new tuberculosis preventive medications are non-inferior to a reference treatment with respect to safety. It is a cluster-randomized adaptive platform trial in which all members of a household are assigned to the same treatment arm, and new regimens are compared against a common control arm across multiple endpoints. To emphasize the applicability of our methodology with a straightforward yet representative trial design, our illustrative example simplifies SSTARLET to a parallel, non-adaptive design in which a single new treatment is compared with the reference arm using one endpoint.

We now introduce general notation for the design of trials with clustered data. We denote the data to be collected in the trial by $\mathcal{D}_{c}$, consisting of $N$ observations from clusters $j = 1, \dots, c$. There are $n_j$ observations from cluster $j$ such that $\sum_{j = 1}^cn_j = N$. As described later, we index the data $\mathcal{D}_{c}$ and consider SSD in terms of the number of clusters $c$. This framework accommodates various contexts where clustered data arise. In cluster-randomized trials, individual observations correspond to distinct participants that are randomized to a given treatment in groups or clusters \citep{murray1998design}. In longitudinal studies, individual observations correspond to repeated measurements from the same participant. 

The data $\mathcal{D}_{c} = \{{\bf{Y}}_{N}, {\bf{A}}_{N}, {\bf{X}}_{N \times p} \}$ may consist of observed outcomes ${\bf{Y}}_{N}$, treatment assignments ${\bf{A}}_{N}$, and $p$ additional covariates ${\bf{X}}_{N \times p}$. For clustered data, $q$ unobserved random effects ${\bf{W}}_{c \times q}$ account for the dependence between outcomes from the same cluster. We assume that these random effects are not shared between observations from different clusters. As a result, we make the standard assumption that outcomes in different clusters are independent. 

To illustrate this notation within our example, the outcomes ${\bf{Y}}_{N}$ indicate whether or not participants experienced a grade 3 to 5 adverse event (AE). There are binary treatment indicators ${\bf{A}}_{N}$ but no additional covariates ${\bf{X}}_{N \times p}$ for this example. We assume the outcomes in this trial come from a Bayesian logistic regression model with cluster-specific intercepts ${\bf{W}}_{c \times 1}$ as random effects. This model is  
\begin{equation*}\label{eq:model}
y_{j, i} \sim \text{BIN}(1, \text{expit}(\eta_{j, i})), ~~\text{where}~~ \eta_{j, i} = \beta_{0} + \beta_{1}A_{j,i} + w_{j},
\end{equation*}
where the subscripts are such that $j$ indexes the cluster and $i$ indexes the observation within the $j^{\text{th}}$ cluster. The $\beta_{0}$ and $\beta_{1}$ parameters represent a common intercept and slope and $w_{j} \sim \mathcal{N}(0, \sigma^2)$ are cluster-specific intercepts. Using general notation, the statistical model for the trial is defined using a set of parameters $\boldsymbol{\theta} \in \boldsymbol{\Theta}$ that may include the distributional parameters for the random effects. The parameters for our example are $\boldsymbol{\theta} = (\beta_0, \beta_1, \sigma^2)$. 

The data associated with the outcomes ${\bf{Y}}_{N}$ are summarized by a scalar estimand. As mentioned earlier, we are interested in the marginal individual-level treatment effect. To facilitate the introduction of our methodology, we denote the estimand as a function of the model parameters: $\delta(\boldsymbol{\theta}) \in \mathbb{R}$. This paper focuses on marginal estimands $\delta(\boldsymbol{\theta})$ that can be expressed as a contrast of population quantities for each treatment. We let $A \in \{0,1\}$ denote the treatment assignment, where $A = 1$ and $A = 0$ respectively represent being randomized to the treatment and control groups. We let $\mu(\boldsymbol{\theta}; A)$ be the population mean, marginalized with respect to the distributions of ${\bf{X}}$ and ${\bf{W}}$, corresponding to the treatment group $A$:
  \begin{equation}\label{eq:est_group}
  \mu(\boldsymbol{\theta}; A) = \int_{{\bf{X}}} \int_{{\bf{W}}}\mu(\boldsymbol{\theta}; A, {\bf{X}}, {\bf{W}})p({\bf{X}})p({\bf{W}})d{\bf{W}}d{\bf{X}}.
\end{equation} 
Using a difference-based contrast without loss of generality, we broadly represent the marginal estimand of interest as
  \begin{equation}\label{eq:estimand}
  \delta(\boldsymbol{\theta}) = \mu(\boldsymbol{\theta}; A = 1) - \mu(\boldsymbol{\theta}; A = 0).
\end{equation} 
For illustration, the estimand for our example is the difference between the marginal AE rates in the treatment and control arms:
\begin{equation}\label{eq:estimand.spec}
\delta(\boldsymbol{\theta}) = \int_{w}\text{expit}(\beta_{0} + \beta_{1} + w)p(w)dw - \int_{w}\text{expit}(\beta_{0} + w)p(w)dw.
\end{equation} 

Lastly, we discuss why our methodology provides the option to marginalize over the distribution of the random effects ${\bf{W}}$. Most cluster-specific models are predicated on the existence of latent risk groups that are indexed by the cluster-specific effects \citep{neuhaus1991comparison}. These latent risk groups may not be balanced for all treatment groups in trials with clustered data, impacting the validity of the trial's conclusions. Since these cluster-specific effects are not observed, it is difficult to determine whether they are balanced between treatment groups. We therefore focus on the case where inference is conducted by treating ${\bf{W}}$ as additional variables that we marginalize over. Nevertheless, the theory and methods we develop here can also be applied to trials with conditional estimands $\delta(\boldsymbol{\theta}; {\bf{X}}, {\bf{W}})$ since conditional estimands are conceptually simpler than marginal ones in regression settings.

  \section{Computational Bottlenecks in Trial Design}\label{sec:methods}

  We now discuss the computational burden associated with assessing design operating characteristics in clinical trials. Design operating characteristics are defined with respect to hypotheses about the estimand(s) of interest. For trials with a single primary estimand, the hypotheses that inform decision making are formulated as
  \begin{equation}\label{eq:hyp}
H_0: \delta(\boldsymbol{\theta}) \notin (\delta_L, \delta_U) ~~~ \text{vs.} ~~~ H_1: \delta(\boldsymbol{\theta}) \in (\delta_L, \delta_U),
\end{equation} 
where $\delta_L$ and $\delta_U$ represent lower and upper interval endpoints. This general notation for $\delta_L$ and $\delta_U$ accommodates hypothesis tests based on superiority, non-inferiority, and practical equivalence \citep{spiegelhalter1994bayesian, spiegelhalter2004bayesian}. In this paper, we focus on trials where the decision rule is defined using the posterior probability that $H_1$ is true. 

The posterior of $\boldsymbol{\theta}$ synthesizes information from the prior distribution for $\boldsymbol{\theta}$ and the data $\mathcal{D}_{c}$. In settings with random effects, Bayesian analysis is facilitated using the joint posterior of $(\boldsymbol{\theta}, {\bf{W}})$ given the data, $\mathcal{D}_{c}$. Samples from the posterior distribution of $(\boldsymbol{\theta}, {\bf{W}})$ give rise to posterior samples from the conditional parameter $\mu(\boldsymbol{\theta}; A, {\bf{X}}, {\bf{W}})$ with respect to fixed patterns for the covariates ${\bf{X}}$ and random effects ${\bf{W}}$. Because the estimands in clinical trials are often non-collapsible \citep{daniel2021making, willard2024covariate}, we marginalize the posterior samples of $\mu(\boldsymbol{\theta}; A, {\bf{X}}, {\bf{W}})$ with respect to the distributions of ${\bf{X}}$ and ${\bf{W}}$ to obtain samples from the posterior distribution of $\mu(\boldsymbol{\theta}; A)$ in (\ref{eq:est_group}). The procedure to implement this marginalization is based on reweighting according to repeated sampling from Dirichlet distributions; its details are described in Appendix B of the online supplement. We use this procedure to approximate the posterior distribution of $\delta(\boldsymbol{\theta})$. From this distribution, we can compute the following posterior probability: 
  \begin{equation}\label{eq:post.prob}
  \tau(\mathcal{D}_{c}) = 
           \Pr(H_{1}~|~\mathcal{D}_{c}). 
\end{equation} 

In this work, we focus on decision rules that reject the null hypothesis if $\tau(\mathcal{D}_{c}) \ge \gamma$, where $\gamma \in [0,1]$ is a decision threshold. The hypotheses in (\ref{eq:hyp}) are formulated such that we support the goal of the trial (e.g., superiority, non-inferiority, or practical equivalence) by rejecting $H_0$. For our illustrative example, the treatment arm is non-inferior to the control if the difference in marginal AE rates defined in (\ref{eq:estimand.spec}) is less than 4\%. To support the trial's goal of non-inferiority, we therefore define the alternative hypothesis as $H_1: \delta(\boldsymbol{\theta}) < 0.04$. That is, $H_1$ is based on an interval such that $\delta_L = -\infty$ and $\delta_U = 0.04$.

We must consider the sampling distribution of $\tau(\mathcal{D}_{c})$ to estimate design operating characteristics. To assess sampling distributions of posterior probabilities via simulation, we define various data generation processes for $\mathcal{D}_{c}$. For each simulation repetition, data are generated according to a fixed parameter value $\boldsymbol{\theta}$. Additional parameters are required to generate the cluster sizes $\{n_j\}_{j = 1}^c$, covariates ${\bf{X}}_{N \times p}$, and potential random effects ${\bf{W}}_{c \times q}$; we assign an auxiliary data-generation process $\boldsymbol{\zeta}$ for this purpose. The probability model $\Psi$ characterizes how $\boldsymbol{\theta}$ values are drawn in each simulation repetition $r = 1, \dots, m$. The probability model $\Psi$ could be viewed as a \emph{design} prior \citep{de2007using,berry2010bayesian,gubbiotti2011bayesian} that differs from the \emph{analysis} prior $p(\boldsymbol{\theta})$. It is standard practice to estimate the sampling distribution of $\tau(\mathcal{D}_{c})$ under $\Psi$ as follows \citep{wang2002simulation, berry2010bayesian}. For each simulation repetition $r$, data $\mathcal{D}_{c, r}$ are generated given $\boldsymbol{\theta}_{r} \sim \Psi$, and $\tau(\mathcal{D}_{c, r})$ is computed. Across $m$ simulation repetitions, the collection of obtained $\{\tau(\mathcal{D}_{c, r})\}_{r = 1}^m$ values estimates the sampling distribution of $\tau(\mathcal{D}_{c})$.

We now define operating characteristics with respect to the model from which $\boldsymbol{\theta}$ values are drawn. For a given model $\Psi$, the probability of supporting $H_1$ is
  \begin{equation}\label{eq:doc}
  \mathbb{E}_{\Psi}[\Pr(\tau(\mathcal{D}_{c}) \ge \gamma~|~\boldsymbol{\theta})] = \int \Pr(\tau(\mathcal{D}_{c}) \ge \gamma~|~\boldsymbol{\theta})\Psi(\boldsymbol{\theta})d\boldsymbol{\theta}.
\end{equation} 
Given our simulation results, the probability in (\ref{eq:doc}) is estimated as 
  \begin{equation}\label{eq:power}
  \dfrac{1}{m}\sum_{r=1}^m \mathbb{I}\left\{\tau(\mathcal{D}_{c}) \ge \gamma\right\},
\end{equation} 
where $\mathcal{D}_{c, r}$ are generated using $\boldsymbol{\theta}_r$ obtained via $\Psi$.  The probability of \emph{correctly} supporting $H_1$ based on the data is $\mathbb{E}_{\Psi_1}[\Pr(\tau(\mathcal{D}_{c}) \ge \gamma~|~\boldsymbol{\theta})]$, where $\Psi_1$ is a probability model such that $H_1$ is true. This probability is estimated using (\ref{eq:power}) when $\mathcal{D}_{c, r}$ are generated using $\boldsymbol{\theta}_r$ obtained via $\Psi_1$. Note that this probability is the classical frequentist power when $\Psi_1$ is degenerate. The term assurance is commonly used if $\Psi_1$ incorporates uncertainty about the parametric assumptions used to generate the data \citep{o2001bayesian}. The probability of \emph{incorrectly} supporting $H_1$ is defined as $\mathbb{E}_{\Psi_0}[\Pr(\tau(\mathcal{D}_{c}) \ge \gamma~|~\boldsymbol{\theta})]$, where $\Psi_0$ is a probability model such that $H_0$ is true. Using $\Psi_0$ instead of $\Psi_1$, this probability can be estimated as in (\ref{eq:power}). When $\Psi_0$ is a degenerate model with parameters corresponding to the null value of the estimand, the probability of incorrectly supporting $H_1$ is the frequentist type I error rate. 

The decision threshold $\gamma$ bounds the probability of incorrectly supporting $H_1$. Asymptotic theory renders $\gamma = 1 - \alpha$ a suitable initial choice to maintain a type I error rate of $\alpha$ \citep{bernardo2009bayesian, golchi2023estimating}. However, this threshold must be tuned in simulation studies, especially for trials with a small or moderate number of clusters. The sample size $N$ is a function of the number of clusters $c$ and the auxiliary data-generation process $\boldsymbol{\zeta}$ that determines the cluster sizes $\{n_j\}_{j=1}^c$. In this paper, we consider SSD in terms of the cluster count $c$ for a given process $\boldsymbol{\zeta}$. The value for $c$ is selected to ensure the trial has a large enough probability of correctly supporting $H_1$. For every value of $c$ considered, we must obtain a collection of $\{\tau(\mathcal{D}_{c, r})\}_{r = 1}^m$ values via simulation to estimate this probability in (\ref{eq:power}). The process to obtain the $\{\tau(\mathcal{D}_{c, r})\}_{r = 1}^m$ values is often computationally intensive. However, we can reduce the computational burden by using previously estimated sampling distributions of $\tau(\mathcal{D}_{c})$ to estimate operating characteristics at new $c$ values. We can use this process to conduct SSD for Bayesian trials with substantially fewer simulation repetitions. We propose such a method for trial design in this paper and begin its development in Section \ref{sec:proxy}. 

\section{A Proxy to the Sampling Distribution}\label{sec:proxy}

To substantiate our SSD procedure proposed in Section \ref{sec:power}, we engage with large-sample theory. Our main theoretical result is derived in Theorem \ref{thm1}; it guarantees that all quantiles of the sampling distribution of the logits of posterior probabilities are approximately linear functions of $c$ for sufficiently large cluster counts. This approximate linearity allows us to expedite simulation-based design. The methods that we present here closely follow the structure of previous work of \citet{hagar2024scalable} for non-clustered data.  

Our main theoretical result is based on a proxy to the  sampling distribution of $\tau(\mathcal{D}_{c})$. These proxies are needed for the theory that underpins our proposed methodology. However, our SSD method does not directly use these proxies and instead estimates true sampling distributions of $\tau(\mathcal{D}_{c})$ by simulating samples $\{\mathcal{D}_{c, r}\}_{r=1}^m$ and approximating posterior probabilities as described in Section \ref{sec:methods}. Thus, although the following technical details provide a theoretical foundation for our method, they are not required to appreciate the practical benefits of our SSD methodology. To define the proxy sampling distribution used in Theorem \ref{thm1}, we must first prove a preliminary theoretical result in Lemma \ref{lem1}. This lemma establishes an asymptotic approximation to the posterior of $\delta(\boldsymbol{\theta})$ that is based on the Bernstein-von Mises (BvM) theorem \citep{vaart1998bvm}. 

We therefore require that the four conditions for the BvM theorem are satisfied to apply our methodology. The first assumption is an independently and identically distributed assumption. We require that independence assumptions are satisfied at the cluster level but not at the observation level. The next two conditions concern the likelihood function and are included in the regularity conditions for the asymptotic normality of the maximum likelihood estimator (MLE) \citep{lehmann1998theory}. For reasons described shortly, our methodology also requires that those regularity conditions are satisfied. The final condition for the BvM theorem concerns the analysis prior $p(\boldsymbol{\theta})$. This prior must be absolutely continuous with positive density in a neighbourhood of the true value for $\boldsymbol{\theta}$. For simulation repetition $r$, this true value of $\boldsymbol{\theta}_{r} \sim \Psi$ is incorporated into Lemma \ref{lem1}. 

\begin{lemma}\label{lem1}
    Assume the conditions described above are satisfied for $\boldsymbol{\theta}_{r} \sim \Psi$. Let $\hat{\delta}^{_{(c)}}_r = \delta(\hat{\boldsymbol{\theta}}^{_{(c)}}_r)$ be the maximum likelihood estimate for $\delta(\boldsymbol{\theta})$ expressed in terms of the number of independent clusters $c$. As $c$ increases, a large-sample approximation to the posterior of $\delta(\boldsymbol{\theta})$ takes the form 
  \begin{equation}\label{eq:bvm}
  \mathcal{N}\left(\hat{\delta}^{_{(c)}}_r, c^{-1}\Lambda(\boldsymbol{\theta}_r)\right),
\end{equation}
  where $\Lambda(\boldsymbol{\theta}_r)$ is related to the Fisher information matrix $\mathcal{I}(\boldsymbol{\theta})$.
\end{lemma}  

In Appendix A.1 of the supplement, we prove Lemma \ref{lem1} and specify the form of the matrix $\Lambda(\boldsymbol{\theta}_r)$. We also carefully justify that Lemma \ref{lem1} holds true for marginal estimands as defined in (\ref{eq:estimand}) when the cluster sizes $\{n_j\}_{j=1}^c$ are drawn according to $\boldsymbol{\zeta}$. Here, we use the approximation in (\ref{eq:bvm}) for theoretical development. The theory in Appendix A.1 additionally demonstrates that the approximate sampling distribution of the MLE $\hat{\delta}^{_{(c)}} ~|~ \boldsymbol{\theta} = \boldsymbol{\theta}_{r}$ is $\mathcal{N}(\delta_r, c^{-1}\Lambda(\boldsymbol{\theta}_r))$ under the required regularity conditions \citep{lehmann1998theory}, where $\delta_r = \delta(\boldsymbol{\theta}_r)$. For theoretical purposes only, a single realization from this normal distribution could be generated using cumulative distribution function (CDF) inversion and a point $u_r \in [0,1]$:
   \begin{equation*}\label{eq:cdf.inv}
\hat{\delta}^{_{(c)}}_{r} = \delta_{r} + \Phi^{-1}(u_{r})\sqrt{\frac{\Lambda(\boldsymbol{\theta}_{r})}{c}},
\end{equation*} 
where $\Phi(\cdot)$ is the standard normal CDF.

Implementing this process with a sequence of $m$ points $\{u_{r}\}_{r = 1}^m \sim \mathcal{U}([0,1])$ simulates a sample from the approximate sampling distribution of $\hat{\delta}^{_{(c)}}$ according to $\Psi$. We substitute this sample $\{ \hat{\delta}^{_{(c)}}_{ r}\}_{r=1}^m$ into the posterior approximation in (\ref{eq:bvm}) to yield samples of posterior probabilities. We approximate the posterior probability that $H_1$ is true -- that is, the posterior probability that $\delta(\boldsymbol{\theta}) \in (\delta_L, \delta_U)$ -- as
      \begin{equation}\label{eq:proxy}
\tau^{_{(c)}}_{r} = 
   \Phi\left(\dfrac{\delta_{U} - \hat{\delta}^{_{(c)}}_{r}}{\sqrt{c^{-1}\Lambda(\boldsymbol{\theta}_r)}}\right) - \Phi\left(\dfrac{\delta_{L} - \hat{\delta}^{_{(c)}}_{r}}{\sqrt{c^{-1}\Lambda(\boldsymbol{\theta}_r)}}\right).
\end{equation} 
The collection of $\{\tau^{_{(c)}}_{r}\}_{r = 1}^m$ values corresponding to $\{u_{r}\}_{r = 1}^m \sim \mathcal{U}([0,1])$ and $\{\boldsymbol{\theta}_{r}\}_{r = 1}^m \sim \Psi$ define our proxy to the sampling distribution of $\tau(\mathcal{D}_{c})$. We acknowledge that this proxy to the sampling distribution of posterior probabilities relies on asymptotic results, and it may differ materially from the true sampling distribution of $\tau(\mathcal{D}_{c})$ for finite $c$. 

The proxy sampling distribution therefore only motivates our theoretical result in Theorem \ref{thm1}. This result guarantees that the logit of $\tau^{_{(c)}}_{r}$ is an approximately linear function of $c$. We later adapt this result to assess the operating characteristics of a trial across a broad range of $c$ values by estimating the true sampling distribution of $\tau(\mathcal{D}_{c})$ at only \emph{two} values of $c$. Each $\tau^{_{(c)}}_{r}$ value in the proxy sampling distribution depends on the value for $\hat{\delta}^{_{(c)}}_{r}$, which depends on the number of clusters $c$, the parameter value $\boldsymbol{\theta}_{r}$, and the point $u_{r}$. We emphasize that $u_{r} \sim \mathcal{U}([0,1])$ is a stochastic conduit for the data and that $\boldsymbol{\theta}_{r} \sim \Psi$ is stochastic when $\Psi$ is nondegenerate. In Theorem \ref{thm1}, we fix both the $\boldsymbol{\theta}_{r}$ and $u_{r}$ values to explore the behavior of $\tau^{_{(c)}}_{r}$ as a deterministic function of $c$.

\begin{theorem}\label{thm1}
    For any $\boldsymbol{\theta}_{r} \sim \Psi$, let the conditions for Lemma \ref{lem1} be satisfied.
    Define $\text{logit}(x) = \log(x) - \log(1-x)$. We consider a given point $u_{r} \in [0,1]$ and distributions for ${\bf{X}}$, ${\bf{W}}$, and the cluster size $n_j$. The function $\tau^{_{(c)}}_{r}$ in (\ref{eq:proxy}) is such that
    $$\lim\limits_{c \rightarrow \infty} \dfrac{d}{dc}~\text{logit}\left(\tau^{_{(c)}}_{r}\right)= (0.5 - \mathbb{I}\{\delta_{r} \notin (\delta_{L}, \delta_{U})\})\hspace*{1pt}\times\min\left\{\dfrac{(\delta_{U} - \delta_r)^2}{\Lambda(\boldsymbol{\theta}_{r})}, \dfrac{(\delta_{L} - \delta_r)^2}{\Lambda(\boldsymbol{\theta}_{r})}\right\} .$$
\end{theorem} 

We prove Theorem \ref{thm1} in Appendix A.2. \citet{hagar2024scalable} considered a simplified theoretical result that did not accommodate dependent data nor account for marginal estimands. The methods proposed in this paper therefore accommodate a broader set of data-driven comparisons in Bayesian trials. We now discuss the practical implications of Theorem \ref{thm1}. The limiting derivative in Theorem \ref{thm1} is a constant that does not depend on $c$. The linear approximation to $l^{_{(c)}}_{r} = \text{logit}(\tau^{_{(c)}}_{r})$ as a function of $c$ is thus a good global approximation for large values of $c$. This linear approximation should be locally suitable for a range of smaller cluster counts. Therefore, the quantiles of the sampling distribution of $l^{_{(c)}}_{r}$ change linearly as a function of $c$ when $\boldsymbol{\theta}_{r}$ is held constant across simulation repetitions. In Section \ref{sec:power}, we exploit and adapt this linear trend in the proxy sampling distribution to flexibly model the logits of $\tau(\mathcal{D}_{c})$ as linear functions of $c$ when independently simulating samples $\mathcal{D}_{c, r}$ according to  $\boldsymbol{\theta}_{r} \sim \Psi$. While the proxy sampling distribution is predicated on asymptotic results, we illustrate the good performance of our SSD procedure with finite cluster counts $c$ in Section \ref{sec:starlet}.

   \section{Sample Size Determination Procedure}\label{sec:power}

   We generalize the results from Theorem \ref{thm1} to develop a procedure for Bayesian SSD that can easily be implemented when estimating posterior probabilities by simulating data $\mathcal{D}_c$. This procedure is described in Algorithm \ref{alg1}. It performs well with a moderate or large number of clusters, and we estimate the sampling distribution of $\tau(\mathcal{D}_c)$ at \emph{only} two values of $c$: $c_0$ and $c_1$. This initial cluster count $c_0$ can be selected based on the anticipated budget for the trial or a less intensive frequentist sample size calculation. In Algorithm \ref{alg1}, we add a subscript to $\mathcal{D}_{c,r}$ to distinguish whether the data are generated according to the model $\Psi_0$ or $\Psi_1$ defined in Section \ref{sec:methods}. That is, $\mathcal{D}_{c,0,r}$ and $\mathcal{D}_{c,1,r}$ represent data generated according to $\Psi_0$ and $\Psi_1$, respectively. The subscript on the number of clusters $c$ is not related to the hypothesis. 
   
   In addition to the choices discussed in Section \ref{sec:methods}, we specify a distribution for the $n_j \times 1$ vector of responses ${\bf{Y}}_j$ at the cluster level, denoted broadly by $p({\bf{Y}}_j; \boldsymbol{\theta})$. We also specify distributions and parameters to generate the cluster sizes $\{n_j\}_{j = 1}^c$, covariates ${\bf{X}}_{N \times p}$, and random effects ${\bf{W}}_{c \times q}$ by way of the auxiliary data-generation process $\boldsymbol{\zeta}$. To implement Algorithm \ref{alg1}, we define criteria for the probabilities correctly and incorrectly concluding $H_1$ is true. Under $\Psi_1$ where $H_1$ is true, we want $\mathbb{E}_{\Psi_1}[\Pr(\tau(\mathcal{D}_{c}) \ge \gamma ~|~\boldsymbol{\theta})] \ge 1 - \beta$, where $\beta$ represents the type II error rate. We want $\mathbb{E}_{\Psi_0}[\Pr(\tau(\mathcal{D}_{c}) \ge \gamma~|~\boldsymbol{\theta})]  \le \alpha$ under $\Psi_0$ where $H_0$ is true. We first list Algorithm \ref{alg1} as pseudocode and then describe the key steps to assist with its implementation.
   

   \begin{algorithm}
\caption{Procedure to Determine the Number of Clusters}
\label{alg1}

\begin{algorithmic}[1]
\Procedure{ClusterSSD}{$p({\bf{Y}}_j; \boldsymbol{\theta})$, $\delta(\cdot)$, $\delta_L$, $\delta_U$, $p(\boldsymbol{\theta})$, $\Psi_0$, $\Psi_1$, $m$, $c_0$, $\boldsymbol{\zeta}$, $\alpha$, $\beta$}
\State  Compute $\{\tau(\mathcal{D}_{c_0, 0, r})\}_{r = 1}^m$ obtained with $\boldsymbol{\theta}_r \sim \Psi_0$. 
\State Choose $\gamma$ to ensure $m^{-1}\sum_{r=1}^m\mathbb{I}\{\tau(\mathcal{D}_{c_0, 0, r}) \ge \gamma\} \le \alpha$.
\State  Compute $\{\tau(\mathcal{D}_{c_0, 1, r})\}_{r = 1}^m$ obtained with $\boldsymbol{\theta}_r \sim \Psi_1$. 
\State If $m^{-1}\sum_{r=1}^m\mathbb{I}\{\tau(\mathcal{D}_{c_0, 1, r})\ge \gamma\} \ge 1-\beta$, choose $c_1 < c_0$. If not, choose $c_1 > c_0$.
\State  Compute $\{\tau(\mathcal{D}_{c_1, 1, r})\}_{r = 1}^m$ obtained with $\boldsymbol{\theta}_r \sim \Psi_1$. 
 \For{$r$ in 1:$m$}
\State Join the $r^{\text{th}}$ order statistics of $\{l(\mathcal{D}_{c_0, 1, k})\}_{k = 1}^m$ and $\{l(\mathcal{D}_{c_1, 1, k})\}_{k = 1}^m$ with a line to obtain $\hat{l}(\mathcal{D}_{c, 1, r})$ \linebreak \hspace*{28pt} for new $c$ values.
 \EndFor
\State Obtain $\{\hat{\tau}(\mathcal{D}_{c, 1, r})\}_{r = 1}^m$ as the inverse logits of the estimates $\{\hat{l}(\mathcal{D}_{c, 1, r})\}_{r = 1}^m$.
\State Find $c_2$, the smallest $c \in \mathbb{Z}^+$ such that $m^{-1}\sum_{r=1}^m\mathbb{I}\{\hat{\tau}(\mathcal{D}_{c, 1, r}) \ge \gamma\} \ge 1-\beta$.
 \State \Return $c_2$ as recommended $c$

\EndProcedure

\end{algorithmic}
\end{algorithm}

In Line 3 of Algorithm \ref{alg1}, we choose a suitable threshold $\gamma$ to ensure the estimate for $\mathbb{E}_{\Psi_0}[\Pr(\tau(\mathcal{D}_{c}) \ge \gamma~|~\boldsymbol{\theta})]$ based on the formula in (\ref{eq:power}) is at most $\alpha$. In Line 5, we choose a second number of clusters $c_1$ at which to estimate the sampling distribution of $\tau(\mathcal{D}_{c})$ under $\Psi_1$. This cluster count could be selected using various strategies. For instance, logits $\{l(\mathcal{D}_{c, 1, r})\}_{r=1}^m$ can be estimated for new $c$ values using lines that pass through the points $\{(c_0, l(\mathcal{D}_{c_0, 1, r}))\}_{r=1}^m$ with the limiting slopes from Theorem \ref{thm1}; the value for $c_1$ could be found as the smallest cluster count such that power (or assurance) based on these estimated logits is at least $1 - \beta$. An alternative strategy is to select $c_0$ and $c_1$ to facilitate visualization of the operating characteristics over a range of relevant cluster counts. 

We emphasize that all posterior probabilities approximated in Lines 2 to 6 of Algorithm \ref{alg1} are obtained by simulating data according to $\Psi_0$ or $\Psi_1$. One strength of our methodology is that it can be integrated with any computational or analytical method to estimate posterior probabilities. Lines 7 to 9 compute logits of these probabilities under $\Psi_1$: $l(\mathcal{D}_{c, 1, r}) = \text{logit}(\tau(\mathcal{D}_{c, 1, r}))$. If a computational method is used to generate posterior samples, we recommend calculating posterior probabilities using a nonparametric kernel density estimate of the posterior distribution so that these logits are finite. We construct linear approximations using these logits in Line 8. We use these linear approximations to estimate logits of posterior probabilities for new values of $c$ as $\hat{l}(\mathcal{D}_{c, 1, r})$. We place a hat over the $l$ here to convey that this logit was estimated using a linear approximation instead of a sample with $c$ clusters. 

Given the linear trend in the proxy sampling distribution quantiles discussed in Section \ref{sec:proxy}, it is reasonable to construct these linear approximations based on order statistics of estimates of the true sampling distributions when the value of the individual estimand $\delta_{r}$ is similar for all $\boldsymbol{\theta}_{r} \sim \Psi_1$. When $\Psi_1$ is nondegenerate, the process in Line 8 can be modified. We instead split the logits of the posterior probabilities for each $c$ value into subgroups based on the order statistics of their $\delta_{r}$ values before constructing the linear approximations. In Line 11, we find the smallest value of $c$ such that our estimate for $\mathbb{E}_{\Psi_1}[\Pr(\tau(\mathcal{D}_{c}) \ge \gamma~|~\boldsymbol{\theta})]$ based on $\{\hat{\tau}(\mathcal{D}_{c, 1, r})\}_{r = 1}^m$ is at least $1-\beta$. 

 We did not estimate the sampling distribution of $\tau(\mathcal{D}_{c_1})$ under $\Psi_0$ in Algorithm \ref{alg1}. We generally consider models $\Psi_0$ that assign all weight to $\boldsymbol{\theta}_{r}$ values such that $\delta(\boldsymbol{\theta}_{r})$ equals $\delta_{L}$ or $\delta_{U}$. For such models $\Psi_0$, large-sample results \citep{bernardo2009bayesian, golchi2023estimating} guarantee that the expected type I error rate for a particular choice of $\gamma$ is roughly constant across a range of large $c$ values. If using a more general model $\Psi_0$, Algorithm \ref{alg1} can be adapted to implement the process in Lines 6 to 10 under $\Psi_0$ to efficiently estimate the sampling distribution of $\tau(\mathcal{D}_{c})$ for new values of $c$. These estimated sampling distributions could be used to choose an optimal threshold $\gamma$ for each cluster count $c$ considered. 

 Lastly, we quantify the impact of simulation variability on the cluster count recommendation by constructing bootstrap confidence intervals for the optimal cluster count. We construct these confidence intervals by sampling $m$ times with replacement from $\{\tau(\mathcal{D}_{c_0, 1, r})\}_{r = 1}^m$ and $\{\tau(\mathcal{D}_{c_1, 1, r})\}_{r = 1}^m$ obtained in Algorithm \ref{alg1}. We obtain a new cluster count recommendation by implementing the process in Lines 7 to 11 of Algorithm \ref{alg1} with the \emph{bootstrap} estimates of the sampling distributions at $c_0$ and $c_1$. This process is repeated $M$ times, and a bootstrap confidence interval for the optimal $c$ is calculated using the percentile method \citep{efron1982jackknife}. The width of this confidence interval can help inform the choice for the number of simulation repetitions $m$. The suitable coverage properties of such confidence intervals were confirmed in \citet{hagar2024scalable}.

Bootstrap confidence intervals for the design operating characteristics at a given value of $c$ can similarly be constructed. For each of the $M$ sets of bootstrap samples, the linear approximations obtained using the process in Lines 7 to 9 of Algorithm \ref{alg1} give rise to a new estimate of the operating characteristics. A bootstrap confidence interval for the operating characteristic can also be calculated using the percentile method. In Section \ref{sec:starlet}, we consider the performance of Algorithm \ref{alg1} and construct bootstrap confidence intervals for our illustrative example.

    \section{Numerical Studies}\label{sec:starlet}

    We now apply the proposed approach to our illustrative example that is inspired by the SSTARLET trial. In this example, our goal is to show that the new tuberculosis preventative medication is non-inferior to the control regimen with respect to safety. In particular, we aim to support the hypothesis $H_1: \delta(\boldsymbol{\theta}) < 0.04$, where $\delta(\boldsymbol{\theta})$ is the difference between the marginal AE rates in the treatment and control arms. Notation for the data, estimand, and analysis model for this example was introduced in Section \ref{sec:ex}. Notation for the hypotheses and posterior probabilities was introduced in Section \ref{sec:methods}.

 Below, we overview the data-generation processes for this example. We consider four design scenarios for this trial that are summarized in Table \ref{tab:scen}. The first three scenarios correspond to non-inferiority under $H_1$ and the fourth corresponds to inferiority under $H_0$. We aim to select a cluster count $c$ such that the power to correctly conclude non-inferiority for the clearly acceptable treatment in scenario 1 is at least $1-\beta = 0.8$. The decision threshold $\gamma$ will be selected to bound the type I error rate for the unacceptable treatment in scenario 4 by $\alpha = 0.025$. The probabilities of concluding non-inferiority in scenarios 2 (acceptable) and 3 (barely acceptable) are considered for sensitivity analysis.

     \begin{table}[ht]
\centering
\caption{Marginal probabilities for the AE rate in various design scenarios}\label{tab:scen}
\begin{tabular}{p{3.35cm} p{1.65cm} p{1.9cm} p{1.9cm} p{1.9cm} p{2.2cm}}
\hline
\multirow{2}{=}{Arm} & \multirow{2}{=}{Reference} & \multicolumn{4}{c}{Experimental Treatment} \\ \cline{3-6} 
 &  & 1: Clearly \newline Acceptable  & 2: \newline Acceptable & 3: Barely \newline Acceptable & 4:  \newline Unacceptable \\ \hline
AE rate & 2\% & 2\% & 3\% & 5\% & 6\% \\
\hline
\end{tabular}
\end{table}

Furthermore, we use three intraclass correlation coefficient (ICC) settings -- low, moderate, and high -- to explore various levels of within-cluster dependence between the AE outcomes. Since we marginalize over cluster-specific random effects in this example, the ICC settings materially impact sample size recommendations. It is therefore important to efficiently consider various ICC settings in trial design. In Appendix C, we thoroughly describe the data generation procedure for all described settings and the diffuse analysis priors used for all parameters.

We first present the simulation results for the low ICC setting. When implementing Algorithm \ref{alg1} with $m = 10^4$ simulation repetitions, an initial cluster count of $c_0 = 100$ was selected based on an anticipated budget for the trial. The probability model $\Psi_0$ for this example corresponds to scenario 4 in Table \ref{tab:scen}. Based on the estimated sampling distribution obtained in Line 2 of Algorithm \ref{alg1}, we chose $\gamma = 0.97$ in Line 3. This threshold was the smallest one rounded to two decimal places that bounded the type I error rate (the probability of an incorrect non-inferiority conclusion) by $\alpha = 0.025$

The probability model $\Psi_1$ for this example corresponds to scenario 1 in Table \ref{tab:scen}. For this $\Psi_1$ model and $c_0 = 100$, the estimated power was 0.7200, which is less than the target probability of $1-\beta = 0.8$. Given that estimate, we next explored a cluster count of $c_1 = 140$ to facilitate assessment of the operating characteristics for $c \in [80, 160]$. The linear approximations in Lines 7 to 9 were constructed using $c_0 = 100$ and $c_1 = 140$. We note that the probability models $\Psi_0$ and $\Psi_1$ for this example do not incorporate uncertainty in the parameter values used to generate the data. These choices for $\Psi_0$ and $\Psi_1$ mirror those used to design SSTARLET, but our methodology accommodates nondegenerate $\Psi$ models as described in Section \ref{sec:power}. Based on the linear approximations in Algorithm \ref{alg1}, the recommended cluster count in Line 11 for the low ICC setting is $c_2 = 115$. A 95\% bootstrap confidence interval for the optimal cluster count obtained using the procedure detailed in Section \ref{sec:power} with $M = 10^4$ bootstrap samples is $[114, 116]$. 

                     \begin{figure}[!tb] \centering 
		\includegraphics[width = \textwidth]{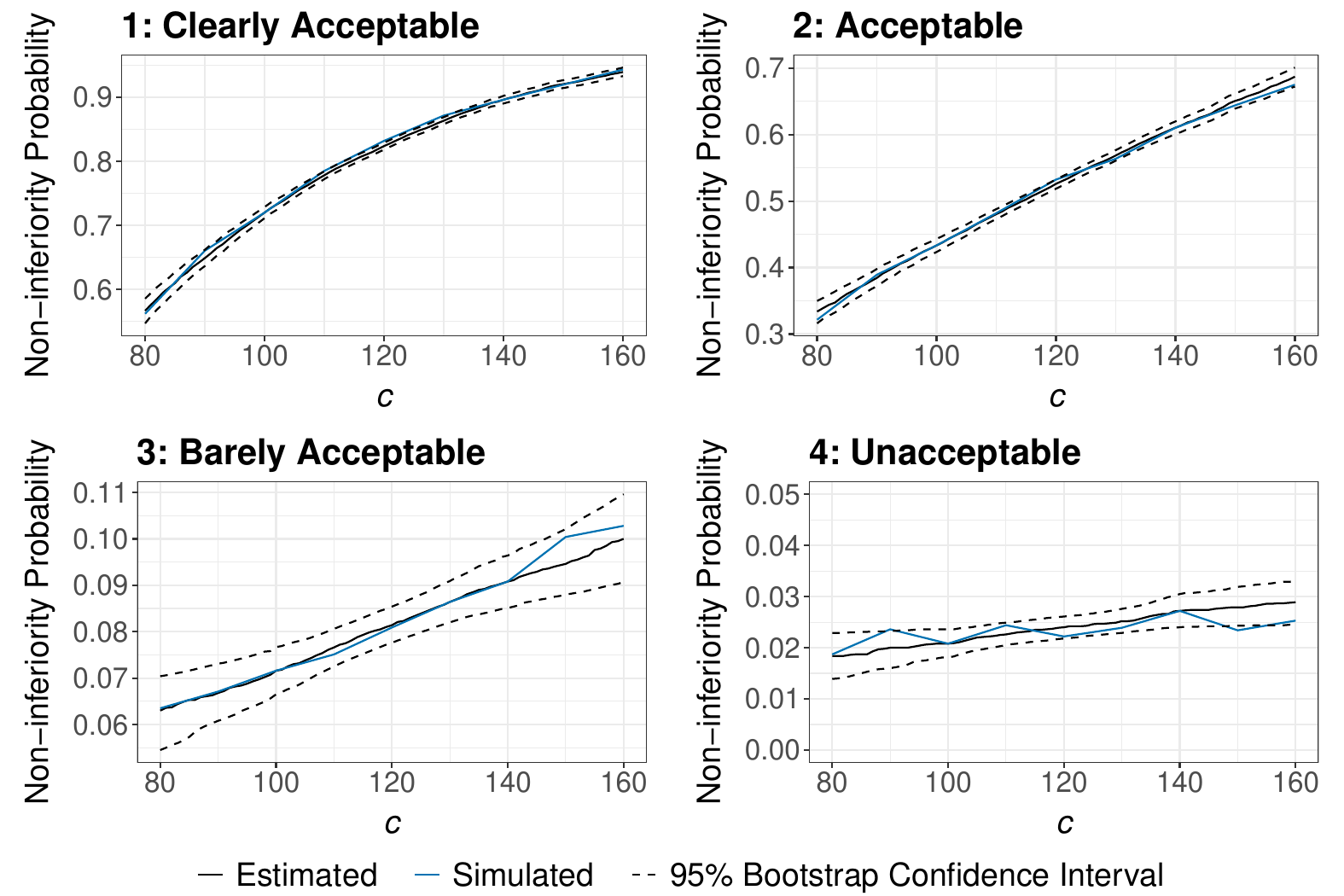} 

		\caption{\label{fig:icc.low} The probability of concluding non-inferiority for the low ICC setting. The black curves are estimated using linear approximations. The dashed curves are pointwise 95\% bootstrap confidence intervals. The blue curves arise from simulating sampling distributions for many $c$ values.} 
	\end{figure}

Figure \ref{fig:icc.low} visualizes the probability of concluding non-inferiority at the final analysis for each scenario in Table \ref{tab:scen} with respect to $c$ for the low ICC setting. This probability corresponds to power for scenarios 1, 2, and 3 and the type I error rate for scenario 4. The black curves were estimated using linear approximations to logits of posterior probabilities at only two cluster counts ($c_0 = 100$ and $c_1=140$) using the process in Lines 4 to 9 of Algorithm \ref{alg1}. The dashed curves represent pointwise 95\% bootstrap confidence intervals for the design operating characteristics obtained using linear approximations with bootstrap samples as described in Section \ref{sec:power}. The blue curves were simulated by generating samples of data to estimate sampling distributions of posterior probabilities at $c = \{80, 90, \dots, 160 \}$ for each setting. 

Although the blue curves are evidently impacted by simulation variability, we use them as approximate surrogates for the true design operating characteristics. We observe good alignment for all four scenarios between the black curves estimated using linear approximations and the blue ones obtained by separately simulating samples. Furthermore, the blue curves are generally contained within the pointwise 95\% bootstrap confidence intervals for the range of cluster counts we consider. However, the black curves are substantially easier to obtain since we need only estimate sampling distributions of posterior probabilities at two values of $c$. 

It took roughly 8 minutes on a high-computing server to estimate each black curve in Figure \ref{fig:icc.low} when approximating each posterior using Markov chain Monte Carlo with only one chain of 500 burnin iterations and 2000 retained draws. While initial simulations used multiple Markov chains to verify convergence, our large-scale simulations employed less intensive settings that achieved reasonable convergence. We considered 9 values of $c$ to simulate each blue curve in Figure \ref{fig:icc.low}, taking approximately 35 minutes using the same computing resources. For standard simulation-based methods, binary search could recommend a cluster count $c \in [1,B]$ by estimating the sampling distribution of $\tau(\mathcal{D}_{c})$ at $\log_2(B)$ values of $c$, where $\log_2(B)$ is generally much greater than two. Unlike standard methods, our approach also allows practitioners to assess design operating characteristics for new values of $c$ without conducting additional simulations.

\begin{figure}[!tb] \centering 
		\includegraphics[width = \textwidth]{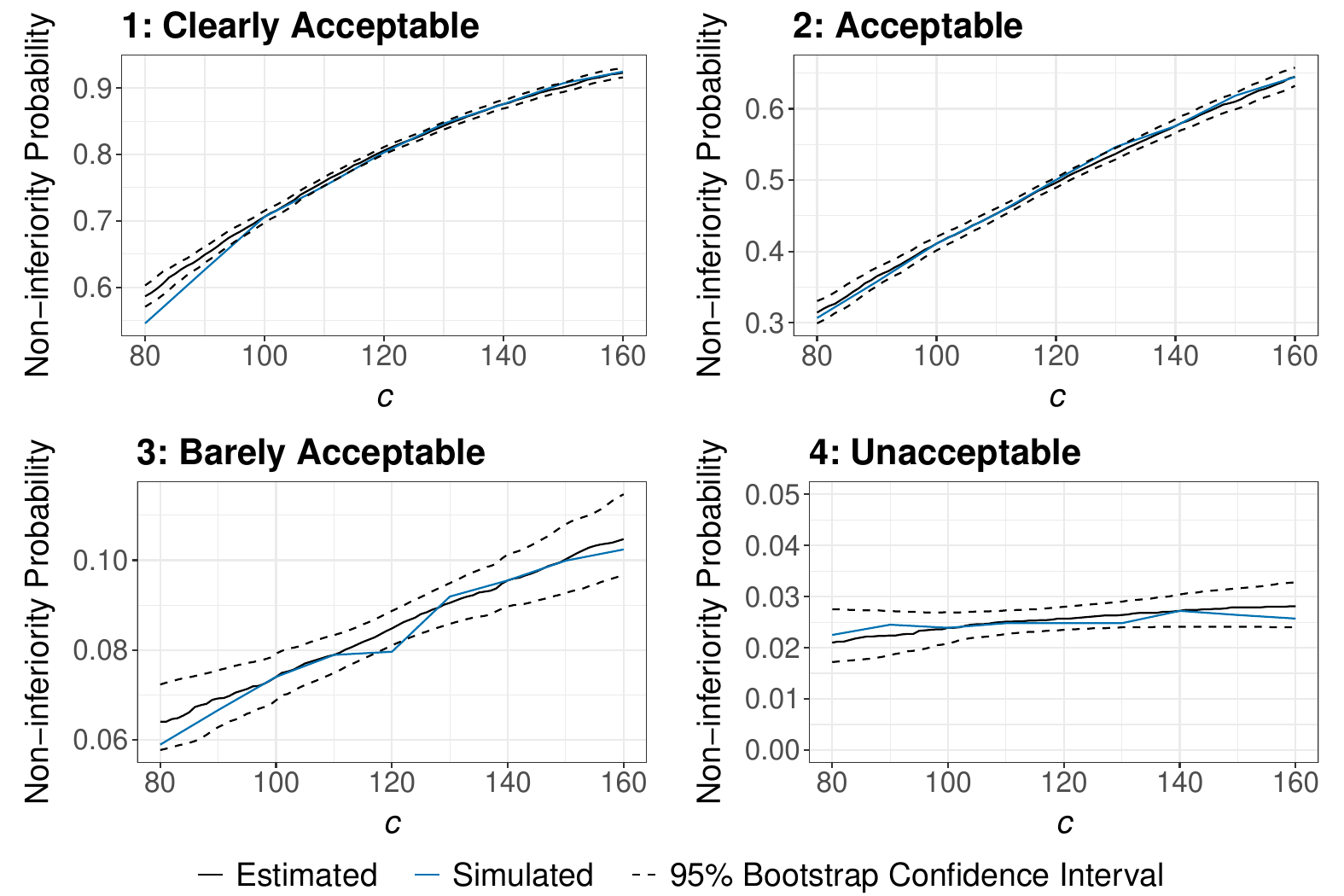}

    		\caption{\label{fig:icc.mod} The probability of concluding non-inferiority for the moderate ICC setting. The black curves are estimated using linear approximations. The dashed curves are pointwise 95\% bootstrap confidence intervals. The blue curves arise from simulating sampling distributions for many $c$ values.} 
	\end{figure}

\begin{figure}[!tb] \centering 
		\includegraphics[width = \textwidth]{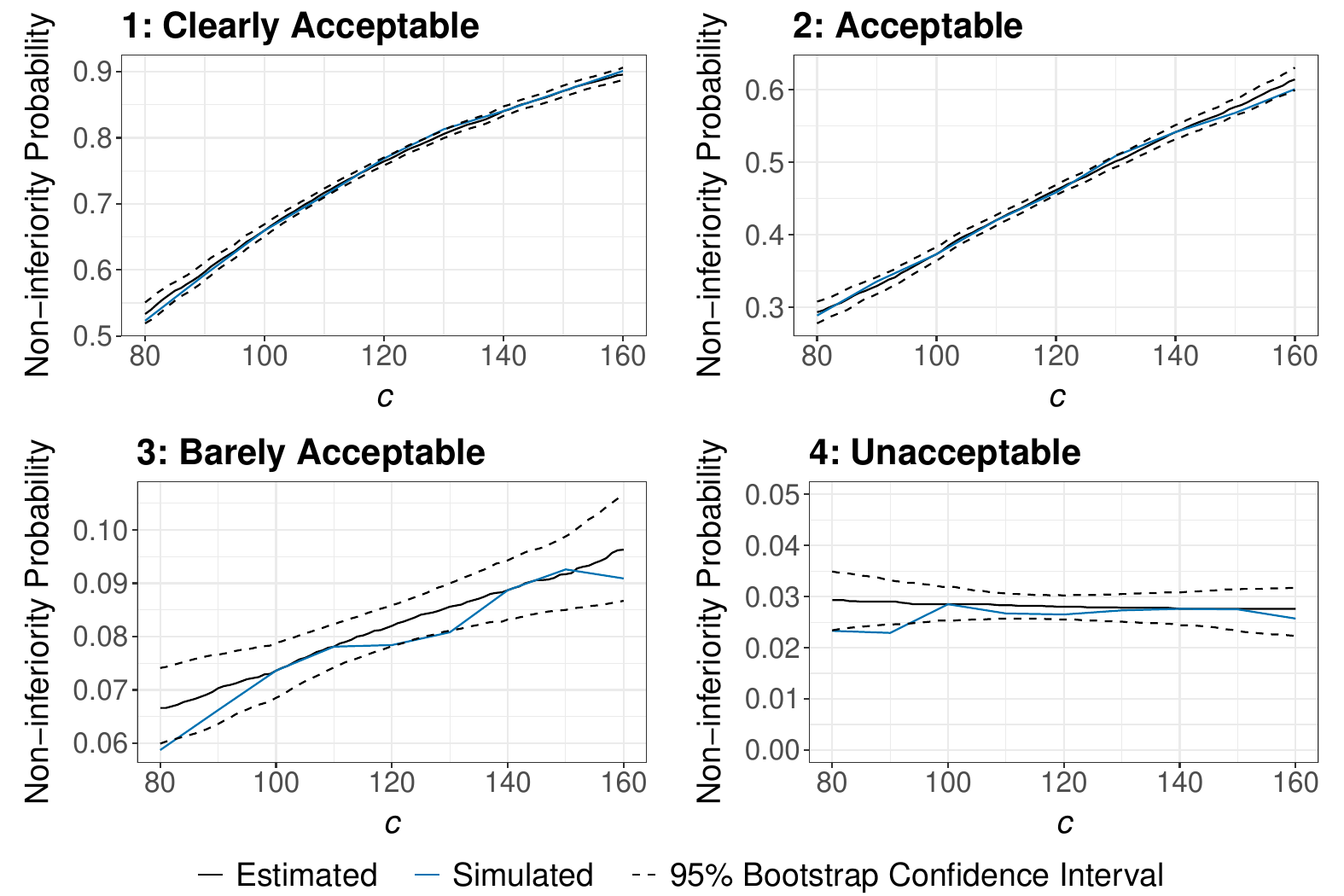} 

		\caption{\label{fig:icc.high} The probability of concluding non-inferiority for the high ICC setting. The black curves are estimated using linear approximations. The dashed curves are pointwise 95\% bootstrap confidence intervals. The blue curves arise from simulating sampling distributions for many $c$ values.} 
	\end{figure}

To facilitate comparison with the low ICC setting, we repeated the process described above with the same decision threshold $\gamma = 0.97$ for the moderate and high ICC settings. Figures \ref{fig:icc.mod} and \ref{fig:icc.high} respectively visualize the probability of concluding non-inferiority for the moderate and high ICC settings. We again observe good alignment between the black curves estimated using our method and the blue ones obtained by traditional simulation. The computational savings associated with using our methodology are similar for all ICC settings.

     Our method recommended a cluster count of $c=119$ for the moderate ICC setting; the 95\% bootstrap confidence interval for the optimal cluster count was [118, 120]. The corresponding cluster count recommendation and confidence interval for the high ICC setting were $c = 129$ and [128, 131]. These results demonstrate that cluster count recommendations are sensitive to the ICC settings when we marginalize over the distributions of the random effects. Thus, it is crucial to explore various ICC settings during trial design, which increases the number of design configurations considered. Our methodology facilitates the efficient assessment of operating characteristics for each configuration.

     To further evaluate the performance of our method, Table \ref{tab:seq.fin} summarizes the estimated probabilities of concluding non-inferiority for the recommended cluster count corresponding to each ICC scenario. The probabilities were first estimated using the linear approximations from Algorithm \ref{alg1} that were anchored at $c_0 = 100$ and $c_1 = 140$. Each pair of rows compares those probabilities with probabilities obtained by estimating the  sampling distribution of $\tau(\mathcal{D}_{c})$ at the recommended cluster count $c$. We observe good alignment between each pair of rows, demonstrating the strong performance of our method. Table \ref{tab:seq.fin} confirms that the desired power of $1-\beta = 0.8$ and type I error rate of $\alpha = 0.025$ are reasonably maintained. 

     \begin{table}[!b]
\centering
\caption{Estimated probabilities of concluding non-inferiority at the recommended cluster counts. These estimates were obtained using linear approximations from Algorithm \ref{alg1} and by simulating confirmatory estimates of sampling distributions.}\label{tab:seq.fin}
\begin{tabular}{
>{\centering\arraybackslash}p{2cm}
>{\centering\arraybackslash}p{1.05cm}
>{\centering\arraybackslash}p{1.65cm}
>{\centering\arraybackslash}p{1.9cm}
>{\centering\arraybackslash}p{1.9cm}
>{\centering\arraybackslash}p{1.9cm}
>{\centering\arraybackslash}p{2.2cm}
}
\hline
\multirow{2}{*}{ICC Setting} & \multirow{2}{*}{$c$} & \multirow{2}{*}{Method} & \multicolumn{4}{c}{Experimental Treatment} \\ \cline{4-7} 
 & &  & \makecell[l]{1: Clearly \\ Acceptable}  & \makecell[l]{2:\\ Acceptable} & \makecell[l]{3: Barely\\ Acceptable} & \makecell[l]{4:\\ Unacceptable} \\ \hline
 \rule{0pt}{2.5ex} \multirow{2}{*}{Low} &  \multirow{2}{*}{115} & Alg. 1 & 0.8020 & 0.5025 & 0.0793 & 0.0235 \\ 
 &  & Simulation & 0.8110 & 0.5118 & 0.0828 & 0.0212 \\ \hline
 \rule{0pt}{2.5ex} \multirow{2}{*}{Moderate} &  \multirow{2}{*}{119} & Alg. 1 & 0.8015 & 0.4923 & 0.0842 & 0.0257 \\ 
 &  & Simulation & 0.8036 & 0.4911 & 0.0811 & 0.0249 \\  \hline
 \rule{0pt}{2.5ex} \multirow{2}{*}{High} &  \multirow{2}{*}{129} & Alg. 1 & 0.8016 & 0.4974 & 0.0852 & 0.0279 \\ 
 &  & Simulation & 0.8078 & 0.4967 & 0.0803 & 0.0286 \\  \hline
\end{tabular}
\end{table}

     We recognize that many cluster-randomized trials do not have as many clusters as our illustrative example based on SSTARLET. In Appendix D of the supplement, we adjust several of the settings for this example to consider the performance of our method with smaller cluster counts. The numerical studies in Appendix D confirm the suitable performance of our method in such settings. 

\section{Extensions for More Complex Designs}\label{sec:ext}

The economical design methodology proposed in this paper for trials with clustered data can be extended to accommodate more complex designs. Several such extensions are needed for the design of the actual SSTARLET trial. For example, the design framework of \citet{hagar2024scalable}, which we build upon here, has previously been extended in \citet{hagar2025sequential} to group sequential designs involving a single estimand and independent observations. In such designs, an alternative hypothesis $H_1$ similar to that in (\ref{eq:hyp}) is considered across $T$ potential stages of the trial. Criteria for the operating characteristics of group sequential designs, including power and the family-wise error rate, are defined with respect to the joint sampling of posterior summaries across these $T$ analyses.  Particularly for adaptive trials that allow early stopping for both efficacy and futility, the operating characteristics require more complex definitions than those we considered in (\ref{eq:doc}) and (\ref{eq:power}).  

To accommodate sequential decision criteria, \citet{hagar2025sequential} developed theory to account for dependence in the joint sampling distribution of posterior and posterior predictive probabilities across trial stages. That theory was substantiated using a proxy sampling distribution based on the BvM theorem and the joint sampling distribution of the MLE arising from the joint canonical distribution \citep{jennison1999group}. This joint sampling distribution of the MLE is a $T$-dimensional analog to the $\mathcal{N}(\delta_r, c^{-1}\Lambda(\boldsymbol{\theta}_r))$ distribution that we considered in Section \ref{sec:proxy}. Combining the results from \citet{hagar2025sequential} with our results here provides an efficient approach to design adaptive cluster trials with a single estimand, where the joint sampling distribution of posterior probabilities is estimated at only two values of $c$. In such an approach, SSD is conducted in terms of the cluster count under the constraint that the proportion of new participants in each trial stage is held fixed, and decision criteria are optimized using the sampling distribution estimates.  

More recently, Hagar and Maleyeff et al.\ (\citeyear{hagar2025platform}) extended the results from \citet{hagar2025sequential} to design platform trials with multiple estimands in non-clustered settings. That work also accounts for (i) fixed delays between reaching interim recruitment thresholds and performing the corresponding analyses and (ii) response-adaptive randomization resulting from dropping treatment arms.  
Each of the $K > 1$ estimands in the trial is associated with a set of hypotheses, giving rise to multiple complementary hypotheses $\{H_{0,k}\}_{k=1}^K$ and $\{H_{1,k}\}_{k=1}^K$. Multiple hypotheses \emph{further} complicate the definition of decision criteria and design operating characteristics. Suppose that all $K$ estimands are considered in each of the $T$ potential stages of the trial. The operating characteristics for platform trials are defined with respect to the joint sampling distribution of posterior probabilities across these $K$ estimands and $T$ analyses. 

One contribution of Hagar and Maleyeff et al.\ (\citeyear{hagar2025platform}) involved deriving new theory to account for dependence in the joint sampling distribution of posterior probabilities across estimands. Their theory leveraged the BvM theorem and the joint sampling distribution of the MLE $\hat{\boldsymbol{\delta}} \in \mathbb{R}^{KT}$. Due to the fixed delays mentioned above, the covariance matrix of this sampling distribution does not take the form $c^{-1}\boldsymbol{\Lambda}(\boldsymbol{\theta}_r)$, where the cluster count (or sample size for independent data) factors out of the denominator. This factorization was required to create our proxy sampling distribution of posterior probabilities in Section \ref{sec:proxy}, so more intricate theory is needed to construct proxy sampling distributions for platform trials and explore linear trends. Nevertheless, the theory in our paper can be combined with a simplified version of the  SSD methodology from Hagar and Maleyeff et al.\ (\citeyear{hagar2025platform}) to design non-platform cluster trials with multiple endpoints.   

 \section{Discussion}\label{sec:disc}

In this paper, we proposed an efficient framework to assess design operating characteristics for Bayesian clinical trials with clustered data. This framework determines the minimum number of clusters required to ensure power (or assurance) is sufficiently large while bounding the expected type I error rate. The computational efficiency of our framework is motivated by considering a proxy for the  sampling distribution of posterior probabilities. We use the behaviour in this large-sample proxy distribution to justify estimating true sampling distributions at only two cluster counts. Our method, therefore, drastically reduces the number of simulation repetitions required to design Bayesian trials. Furthermore, we repurpose our estimates of the sampling distribution to construct bootstrap confidence intervals that quantify the impact of simulation variability on the cluster count recommendation. Our methodology can be broadly used in clinical settings where large-sample regularity conditions are satisfied – including settings that do not require marginalization via Bayesian G-computation.

Our design methodology efficiently recommends the number of clusters $c$ for trials with clustered data. However, we acknowledge that the cluster count is only one aspect of the total sample size in cluster trials. We assign a  (potentially degenerate) probability distribution for the cluster sizes $\{n_j\}_{j=1}^c$ by specifying the auxiliary data-generation process $\boldsymbol{\zeta}$. Our recommendations for $c$ thus depend on these choices. To consider sample size recommendations with different cluster sizes, we must reimplement our algorithm. Additional simulations would also be required to consider different cluster sizes in a standard simulation-based design framework. Furthermore, we emphasize that the asymptotic results that substantiate our methods apply when the number of clusters $c \rightarrow \infty$. This theory does not apply when there are a small number of clusters with large cluster sizes.

The SSTARLET trial that inspired our illustrative example provides insights for future research described in Section \ref{sec:ext}. Moreover, our example reflects that clinicians often want to consider the trial operating characteristics when parameters $\boldsymbol{\theta}$ are generated according to various probability models $\Psi_1$ and $\Psi_0$. We already efficiently explore the sample size space, but analogs to Theorem \ref{thm1} that enable efficient consideration of the $\Psi_1$-space and $\Psi_0$-space for a given cluster count would be useful. Lastly, the use of dynamic borrowing methods to incorporate information from prior trials has become increasingly common. Depending on the borrowing method used, the conditions for the BvM theorem may not be satisfied. Further research is needed to clarify when these conditions are met and develop extensions for scenarios where the conditions are not satisfied.

 \section*{Supplementary Material}
These materials include proofs of Lemma \ref{lem1} and Theorem \ref{thm1}, as well as additional context for the example in Section \ref{sec:starlet} and additional numerical studies. The code to conduct the numerical studies in the paper is available online: \url{https://github.com/lmhagar/ClusterDOCs}.

\section*{Acknowledgements}
The authors would like to thank Dr. Dick Menzies for his comments regarding the SSTARLET trial that inspired the illustrative example for this paper.

	\section*{Funding}
 
 Luke Hagar acknowledges the support of a postdoctoral fellowship from the Natural Sciences and Engineering Research Council of Canada (NSERC). Shirin Golchi acknowledges support from NSERC, Canadian Institute for Statistical Sciences (CANSSI), and Fonds de recherche du Québec - Santé (FRQS).
	


\bibliographystyle{chicago}


\end{document}